# Soft Magnetic Properties of Ultra-Strong and Nanocrystalline Pearlitic Wires

**Stefan Wurster [1,*], Martin Stückler [1], Lukas Weissitsch [1], Heinz Krenn [2], Anton Hohenwarter [3] Reinhard Pippan [1] and Andrea Bachmaier [1]**

[1] Erich Schmid Institute of Materials Science of the Austrian Academy of Sciences, Jahnstrasse 12, 8700 Leoben, Austria;

[2] Institute of Physics, University of Graz, Universitätsplatz 5, 8010 Graz, Austria;

[3] Department of Materials Physics, Montanuniversität Leoben, Jahnstrasse 12, 8700 Leoben, Austria;

* Correspondence: stefan.wurster@oeaw.ac.at

Abstract

The paper describes the capability of magnetic softening of a coarse-grained bulk material by a severe deformation technique. Connecting the microstructure with magnetic properties, the coercive field decreases dramatically for grains smaller than the magnetic exchange length. This makes the investigation of soft magnetic properties of severely drawn pearlitic wires very interesting. With the help of the starting two-phase microstructure, it is possible to substantially refine the material, which allows the investigation of magnetic properties for nanocrystalline bulk material. Compared to the coarse-grained initial, pearlitic state, the coercivities of the highly deformed wires decrease while the saturation magnetization values increase—even beyond the value expectable from the individual constituents. The lowest coercivity in the drawn state is found to be 520 A m$^{-1}$ for a wire of 24-µm thickness and an annealing treatment has a further positive effect on it. The decreasing coercivity is discussed in the framework of two opposing models: grain refinement on the one hand and dissolution of cementite on the other hand. Auxiliary measurements give a clear indication for the latter model, delivering a sufficient description of the observed evolution of magnetic properties.

**Keywords:** nanocrystalline metal; pearlitic steel; wire drawing; ferromagnetic material; coercivity

## 1. Introduction

Severely drawn pearlitic steels have drawn the attention of scientists as well as engineers due to their exceptional high strength [1]. Going along with the mechanical characterization of this interesting material, substantial efforts regarding microstructural characterization after application of high drawing strains were undertaken [2–6]. Therein, the nanocrystalline microstructure, the dissolution of cementite ($Fe_3C$), and a small tetragonal distortion ($c/a < 1.01$) of the remaining ferrite were revealed. Accompanying the nanocrystalline microstructure (see Figure 1 a,b), an exceptionally high strength of almost 7 GPa was found for wires subjected to a true drawing strain " of up to 6.5 [3,7]. This is the highest strength ever measured for a material produced by metal forming techniques. Furthermore, Li et al. [7] used atom probe tomography (APT) to show that subgrain sizes of 10 nm and below can be found in wires of such a degree of deformation, which is the smallest grain size produced by conventional metal forming techniques. Furthermore, the dissolution of cementite by severe wire drawing using APT is described: $Fe_3C$ is dissolved and carbon is mechanically alloyed into the ferrite but also decorating the grain boundaries. Not only the mechanical properties but also physical properties are affected by the substantial microstructural refinement. It is well known that the microstructure of a ferromagnetic material has a strong influence on the magnetic properties. Consequently, if one alters the microstructure (e.g., by refining the grain size D) in a controlled manner, the coercive field strength $H_c$ can be modified correspondingly. Herzer showed for very small grains from the so-called random anisotropy regime, where the magnetic exchange length $L_{ex}$ covers many grains, that coercivity collapses with $D^6$ for decreasing grain sizes [8,9]. From the seminal work on FeCuNbSiB it is well known that adequate annealing treatments of rapidly quenched, amorphous materials precipitate very small grains enabling the tuning of coercivity [8]. This process can be regarded as a "bottom-up" process, as small crystallites are generated from an amorphous phase. In contrast to this bottom-up process, strong or severe plastic deformation (SPD) methods can be regarded as "top-down" processes, where microstructural changes of coarse materials result in ultra-fine grained or nanocrystalline microstructures resulting in large changes of not only mechanical but also magnetic properties. Consequently, the magnetic properties of pearlitic wires experiencing high-drawing strain, featuring smallest microstructural sizes, are of large interest. The following questions are going to be answered: Does the coercivity further decrease for very high applied strains? Does the random anisotropy model explain the decreasing coercivities with increasing drawing strains, or can the explanation also be found in other microstructural changes?

## 2. Materials and Methods

The hypereutectoid pearlitic steel containing 0.98 wt.% C was initially provided in the shape of a wire with a diameter $d_{start}$ of 540 µm featuring an average grain size of 23 µm, measured by electron backscatter diffraction. Upon severe wire drawing, thin wires with diameters of a few tens of micrometres were produced. Besides the non-deformed state ($\varepsilon = 0$ for $d_{start} = 540$ µm), thin wires experiencing true drawing strains of 5.42 ($d_{finish} = 36$ µm) and 6.23 ($d_{finish} = 24$ µm) were investigated regarding magnetic properties such as coercivity and saturation magnetization.

A superconducting quantum interference device (SQUID, Quantum Design MPMS XL7), a very sensitive magnetometer for detecting tiny changes of magnetic moments in magnetic fields, was used to measure the volume saturation magnetization at very high fields up to 7 T and the variations of coercivity with changing temperature. Checking reproducibility, the saturation magnetization of two wires of 36-µm thickness was determined. SQUID measurements yield the total magnetic moment of the whole sample volume and for an accurate determination of the saturation magnetization, a normalization with respect to the sample volume has to be made. For the determination of the volume of the thickest wire, a conventional laboratory scale (Sartorius Secura225D-1S, Sartorius, Gottingen, Germany) and using the known mass density of pearlite (7845 kg m$^{-3}$ [10]) provides sufficient accuracy. However, it is more difficult to determine the volume of short sections of very thin wires (24 and 36 µm). The sample volume was determined with a confocal laser scanning microscope (Olympus LEXT OLS 4100, Olympus, Tokyo, Japan). It scans along the complete sample and measures the sample height at each point. Under the assumption of cylindricity, the surface, and thus the volume of the wire can be precisely reconstructed. As an example, a small section is provided in Figure 1c. The surface profiles of the upper halves of the wires were fitted by a sequence of circles of varying diameter to determine the wires' diameter as a function of length. By integrating along the wire axis, the probed sample volume is obtained to a higher accuracy compared to only using the nominal diameter and the length of the wire. To give an idea on the probed sample sizes, the volume of the thinnest wire (see Figure 1d) was determined to be 8.9*10$^{-4}$ mm$^3$.

This extremely small sample volume, albeit of ferromagnetic nature, generates a very small SQUID-signal. As the SQUID should detect only the tiny magnetic moment of the wire, it is necessary to compensate any predominant and inhomogeneous background, e.g., of the sample holder on which the sample is mounted. Following the idea of Topolovec et al., long adhesive tapes were used to fix the sample within the SQUID [11]. These tapes extend beyond the scan range of the second order gradiometer coils of the SQUID flux-transformer. Thus, it cancels the magnetic background signal, but leaves only the signal of the localized wire sample.

The room temperature coercivities of the wires were determined using a vibrating sample magnetometer (VSM, LakeShore 7404, LakeShore Cryotronics, Westerville, OH, USA). Due to the soft magnetic nature of the material, these measurements include the correction of the results for the remanence of the pole pieces, using a paramagnetic Pd standard, which was measured under the same conditions.

For all magnetic measurements (SQUID and VSM), the wires were brought into the magnetic field with the long axis of the wire, being some millimetres of length, aligned with the magnetic field. Due to the large aspect ratio (specimen length/diameter), even for the thickest wire, the influence of shape anisotropy [12] can be considered to be small and almost identical for the measured magnetic properties.

Scanning electron microscopy (Figure 1d) was performed with a LEO1525 (Zeiss).

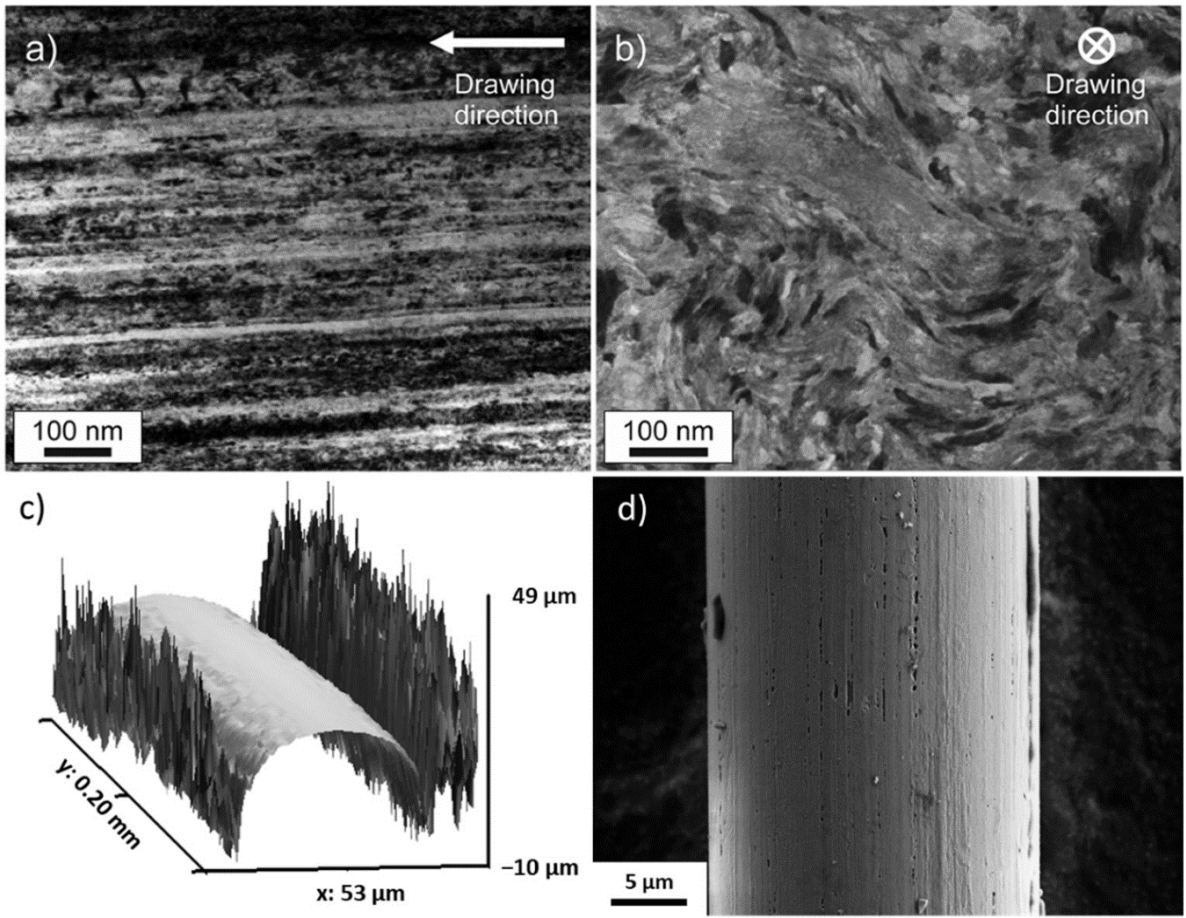

**Figure 1.** (a) Microstructure of the wire with a diameter of 24 µm. Viewing direction is perpendicular to drawing direction (b) Viewing direction is parallel to drawing direction. The typical aligned and elongated (a) and curled (b) microstructure is visible. (a,b) are taken from [13], licensed under Creative Commons Attribution 4.0. (c) 3D-reconstruction of a small section of the wire's surface. Gwyddion 2.53 was used to depict the confocal laser scanning microscopy results [14]. (d) Scanning electron microscopy image of a part of the 24-µm wire.

### 3. Results and Discussion

#### 3.1. Saturation Magnetization

Combining the measured volume and the magnetic moment from SQUID measurements, the volume saturation magnetization can be calculated and the results for all investigated wires are presented in Figure 2. The large increase in magnetization after wire drawing can be explained by the dissolution of cementite. In Figure 2, it can be seen that the saturation magnetization of all thin wires is markedly higher in comparison to the pristine one (540 µm). The magnetization at the highest applied magnetic field of 7 T is about $1.75 \times 10^6$ A m$^{-1}$ for two of the thin wires, while it is slightly larger for one of the two 36-µm wires. The difference, however, can be regarded as a minute one since an uncertainty of only

~2% of the wire's diameter could explain the deviation. In contrast, a much lower value (1.57 ± 0.01) x $10^6$ A m$^{-1}$ was found for the cementite-containing, non-deformed wire.

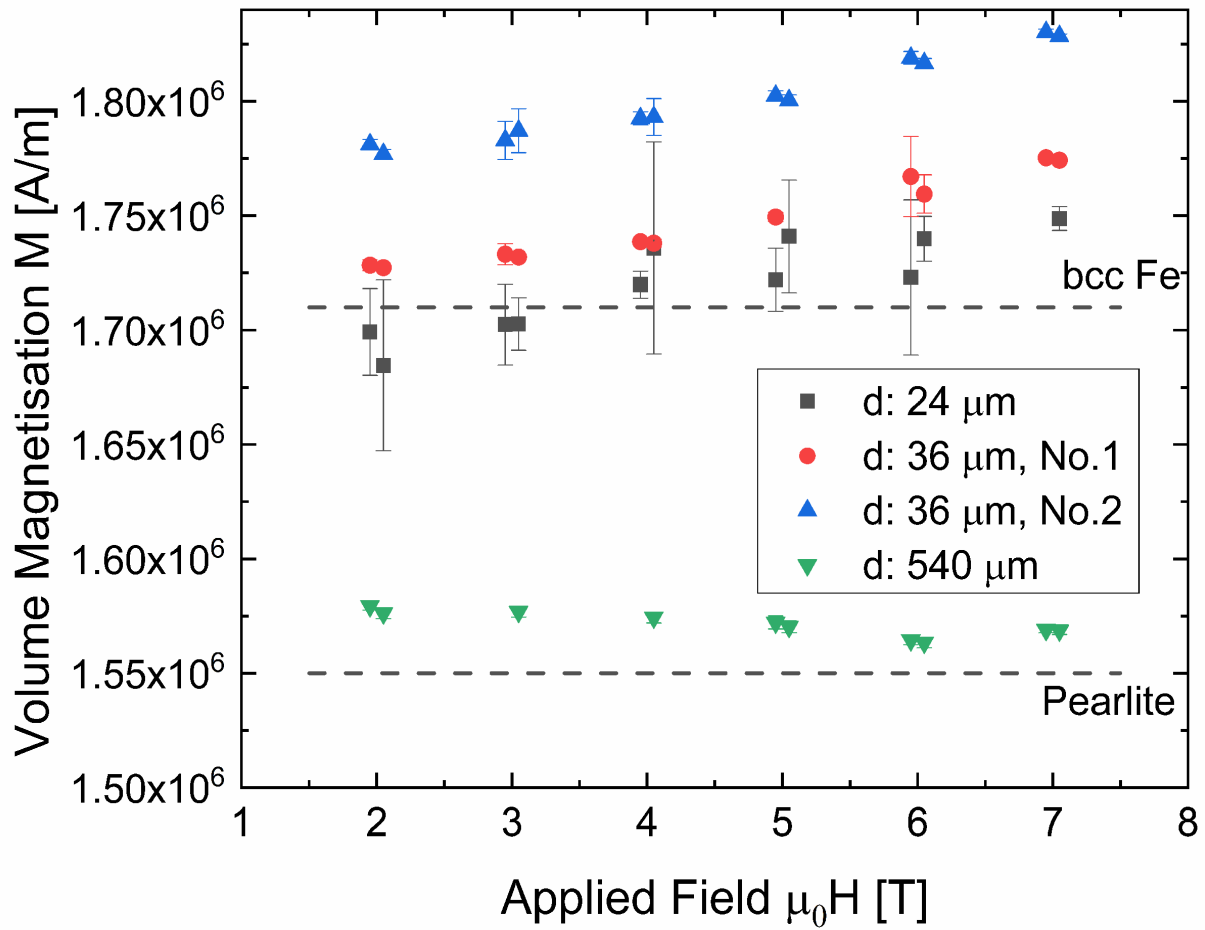

**Figure 2.** Volume saturation magnetization of all investigated wires, measured at applied fields > 2 T. For visualization the repeatability of the measurements, the x-values are slightly shifted for repeated measurements. The error bars only consider the uncertainties of the SQUID measurement, but not the one of the calculated sample volume. The values of bulk bcc Fe and pearlite [15–17] are marked with dashed lines.

For comparison, the value 1.71 x $10^6$ A m$^{-1}$ (2.15 T [15]) for pure body-centered cubic (bcc) Fe is somewhat below all thin wires' magnetization values. Following the idea of Gorkunov et al. [18], one explanation of the increased values of the deformed wires - even in comparison to pure Fe is an increased magnetic moment of Fe atoms with C in supersaturated solid solution. Medvedeva et al. [19] found a magnetic moment of Fe atoms of 2.35 to 2.45 $\mu_B$ when including 1 C-atom in a matrix of 16 Fe-atoms. This calculated magnetic moment is considerably higher than the value for pure bcc Fe (2.17 $\mu_B$ [15]). Cadeville et al. [20] found an increase of 0.02 $\mu_B$ per at% C for splat-quenched Fe–C when C is in supersaturated solid solution. Considering the increase of 0.02 $\mu_B$ per at% C and all available C atoms of the wire to be involved, this would lead to the expected saturation magnetization close to 1.74 x $10^6$ A m$^{-1}$. This coincides very well with the measured values.

In contrast to the thin wires, the saturation magnetization of the thick wire agrees with the value of pure pearlite. Assuming that 0.77 wt.% C (=3.5 at%) are involved in $Fe_3C$ formation, 10.5 at% of Fe are included in the cementite and the rest of Fe forms the ferritic phase. Cementite is ferromagnetic at room temperature. Using the room temperature saturation magnetization of cementite, ~62 emu $g^{-1}$ [16,17] or 0.48 x $10^6$ A $m^{-1}$, together with the above mentioned value for pure Fe and the individual contents of Fe and $Fe_3C$, a saturation magnetization of 1.55 x $10^6$ A $m^{-1}$ for pearlitic material can be calculated, using a simple rule of mixture. The difference between calculated and measured value is below 2%, see Figure 2.

An advantage of pearlitic steels is their chemical composition. An important aspect of the formation of soft magnetic materials is maintaining a high saturation magnetization $J_s$, which can easiest be achieved with high Fe-contents. When using the bottom-up process, several other - very often non-magnetic - elements have to be used to accomplish the goal of formation of nanocrystalline alloys from e.g., amorphous melt-spun ribbons. Suzuki et al. [21] found a correlation between saturation magnetization and additive content, resulting in a change of saturation magnetization by -0.05 T/wt.%$_{Additives}$, independent of the type of the additive. Here it was demonstrated that pearlitic steel even increases its saturation magnetization upon severe deformation with the very small amount of C (in wt.%) being of no relevance in regard of the above-mentioned decrease of $J_s$ with additive content.

In short, while the thick, non-deformed wire shows the expected pearlitic behaviour in saturation magnetization, all thin wires show values near or even above pure Fe. The slight surplus in saturation magnetization is possibly due to C in supersaturated solid solution, which is typically present in the severely deformed wires [6,7].

### 3.2. Coercivity

For an ideal soft magnetic material, the coercivity should tend towards zero, easing the magnetization process and reducing the hysteretic heat losses for certain applications. The coercivity does not necessarily increase with decreasing grain size, as it is the case for "coarse" grained (D > ~100 nm) materials [22]. For very small grains from the so-called random anisotropy regime, where $L_{ex}$ covers many grains, coercivity collapses with $D^6$ [8,9]. The relevant length scale, $L_{ex}$, is derived from a competition between magnetic exchange and magneto-crystalline anisotropy energy $K_1$ and it is given by:

$$L_{ex} = \sqrt{\frac{A}{\beta K_1}} \qquad (1)$$

with A being the exchange stiffness constant (Fe: 20.7 pJ/m[15], $K_1$ = 45,000–46,800 J/$m^3$ [23–25]. b takes into account the symmetry of the crystal, considering the distribution of the random anisotropy axes, which results in b = 0.4 for cubic crystals [26]. The opposing effect of D on the coercivity for D < $L_{ex}$ and D > $L_{ex}$ are described by Equations (2) and (3) [8].

$$H_c = p_c \frac{\langle K \rangle}{J_s} \qquad \text{for D < L}_{ex} \qquad (2)$$

$$H_c = p_c \frac{\sqrt{A K_1}}{J_s D} \qquad \text{for D > L}_{ex} \qquad (3)$$

$J_s$ is the magnetic saturation polarization in [T] (Fe: 2.15 T [15]), $p_c$ is a dimensionless pre-factor close to unity [8] and the average anisotropy constant $\langle K \rangle = \beta K_1^4 D^6 A^{-3}$. Using $L_{ex} = D$, the threshold grain size for Fe, giving highest $H_c$ based on above equations, is found to be close to 34 nm. Applying the given values of pure Fe using above equations, a grain size of 10 nm is needed to achieve a soft material with a coercivity of 100 A m$^{-1}$ and about 5 nm for a very soft material with 1 A m$^{-1}$.

Focusing on pearlitic steel, References [18,27–29] describe the interplay of coercivity and plastic deformation. In [18], the changes of the magnetic properties of patented steels containing 0.24 and 0.70 wt.% C are discussed. An increasing saturation magnetization with increasing true strain due to cementite dissolution was found. The coercivity increases at small drawing strains, which is followed by a constant regime for the material containing 0.24 wt.% ($H_{c,const}$ ~1200 A m$^{-1}$). Contrarily, for the 0.70 wt.% material a peak at low strains (ε ~0.4, $H_c$ ~1600 A m$^{-1}$) is followed by a drop in coercivity, reaching about 1200 A m$^{-1}$ at a true strain of ~3.5. The decreasing coercivity is explained by a combination of volume reduction of cementite as well as rotation and fragmentation of cementite lamellae. These arguments are complemented by Ul'yanov et al. who found a low coercivity state ($H_c$ ~8000 A m$^{-1}$) of cementite after strong deformation using ball milling [27].

The measured coercivities $H_c$ of the thin wires, as determined by VSM, are smaller (24 μm: 520 A m$^{-1}$, 36 μm: 610 A m$^{-1}$) than that of the thick wire (540 μm: 1510 A m$^{-1}$). The coercivity of the pristine state is in agreement with results of both, a cold drawn, 0.8%C steel "music wire" ($H_c$ ~1590 A m$^{-1}$) [30] as well as with a steel containing slightly less C: 0.7%, found by Gorkunov et al., with $H_c$ ~1350 A m$^{-1}$ [18].

In a previous work by our group, a strong, beneficial influence of a low temperature annealing treatment (150 °C, 1 h) on the coercivity of severe plastically deformed materials was found [31]. For different Cu-based materials with dilute Fe contents, the drop in coercivity upon annealing was up to 76%. Thus, there is a chance to further enhance the soft magnetic properties by annealing, but this treatment has to be performed in a careful way, since according to Equation (2) a slight increase in grain size already has a huge effect on coercivity. In [32], it was stated that after annealing at 150 °C for 30 min, the strongly deformed wires retain their strength and consequently their refined microstructure. This treatment was chosen for the 36-μm wire. VSM measurement shows that the coercivity drops by 76 A m$^{-1}$ or 12% due to this annealing treatment.

If one strictly connects the grain size with coercivity, the magnetic measurements would indicate that there is hardly any further refinement of grain size when drawing the wire from 36 to 24 μm in diameter, albeit the reduction in diameter constitutes a substantial degree of further deformation. Along with the impart of additional strain, Li et al., observed a further strengthening of the material for increasing the strain from 5.1 ($\sigma_{max}$ ~6.2 GPa) to 6.52 ($\sigma_{max}$ ~6.8 GPa) [7]. If one were to consider that the strength of the material can only be influenced by the grain size, a contradiction arises between the magnetic measurements and the mechanical results. However, in the literature it can be frequently found that the strength is described as a combination of several contributing microstructural features such as dislocation density and extent of carbon dissolution, e.g., shown by Zhang et al. for wires of drawing strains up to 5.4 [33]. The magnetic measurements support this approach to explain the strength of supersaturated nanostructures.

Due to the findings of microstructurally refined and magnetically soft materials, Equation (2) is applied to draw the graph in Figure 3. It demonstrates that the combination of subgrain sizes found in literature [7] and measured coercivities of the thin wires is close to the result of Equation (2). As already mentioned, Li et al. found a size below 10 nm for equiaxed subgrains in the transverse cross section of the wire, which was subjected to $\varepsilon$ = 6.52. It is important to note that the grain size is anisotropic with the grains being elongated along the drawing direction [7]. The larger effective grain size might be the explanation for the deviation of measured data from the trend line.

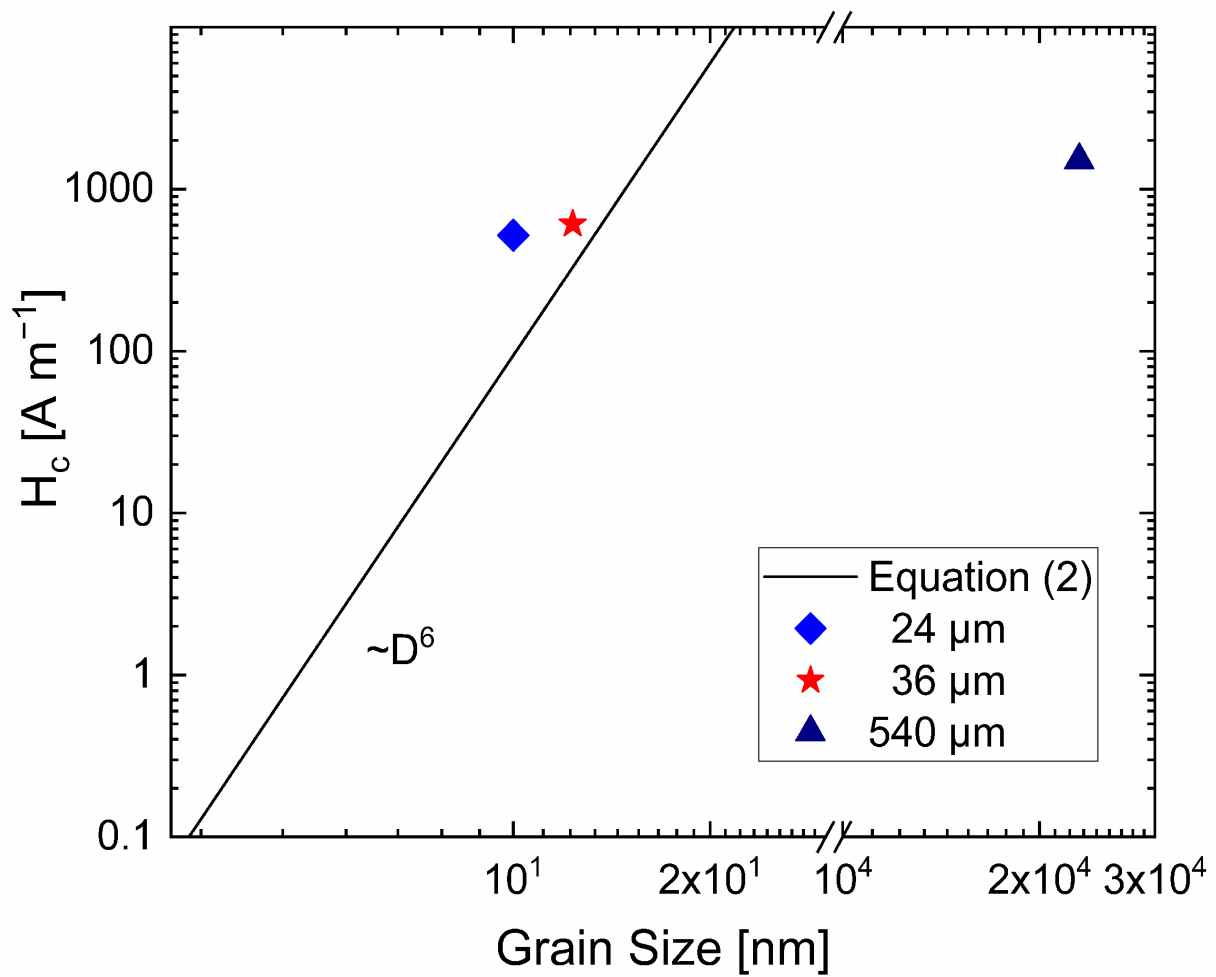

**Figure 3.** Coercivity as a function of small grain sizes, using Equation (2) and the parameters for pure bcc Fe mentioned in the text, compared with the coercivity of thin wires. For the thinnest wire, the subgrain size of 10 nm [7] is taken as the grain size; for the 36-μm wire, the grain size is enlarged by 2 x ln(36/24) to 12.3 nm. The 540 μm wire with a grain size of 23 μm, does not follow this trend.

Regarding the thin wires, there is a small tetragonal distortion (c/a < 1.01) of the bcc crystal present in the severe plastically deformed material [34,35] and furthermore, it was shown that there is a clear difference in the hysteretic behaviour between bcc and tetragonally distorted Fe nanoparticles (c/a = 1.23) [36]. However, due to the much smaller tetragonal distortion found in pearlitic wires, the usage of the magnetic parameters for bcc Fe is justified.

Another proof for the dissolution of cementite to a large extent in the severely drawn wires is provided by temperature dependent measurements of coercivity. Using the SQUID, the coercivity as a function of temperature was measured for a 36-µm wire, which provides - besides the saturation magnetization - another strong hint for the predominance of the bcc Fe phase, but not of the cementite phase. This measurement involves recoil loops of going to the highest fields, saturating the sample and then accurately measuring the coercivity applying small fields. These loops are repeated in temperature steps of 20 K. No Pd-standard was available at that time for SQUID measurements, thus, the measurements of the coercivity always include a systematic offset due to the residual magnetic fields of the superconducting magnet [37]. To approximate the real values, the coercivity of both measurements (VSM, SQUID) at the highest temperatures are subtracted to calculate the offset, which is used to compensate for the systematic offset in SQUID measurements. When applying the same sequence for all measurements, it can be safely assumed that this systematic error (i.e., residual fields) is constant and the measurement can be corrected for the coercivity determined with VSM. Considering the grain size of the 36-µm wire, assumed to be 12.3 nm (see Figure 3), temperature dependent coercivities are calculated, based on temperature dependent magneto-crystalline anisotropies $K_1$ taken from literature [23–25]. A comparison of the calculated and measured values is given in Figure 4. There is an offset of measured data compared to literature data, which is easily explainable by the unknown actual grain size and its strong dependency ($D^6$) on coercivity. However, the trend in measured coercivities perfectly follows the ones in literature. While the magneto-crystalline anisotropy only drops by a bit more than 10% for bcc-Fe [23–25], the drop for $Fe_3C$ would be much larger, about 50% [38]. Taking these results and the values of the volume saturation magnetization, the deformed wires behave magnetically like bcc Fe.

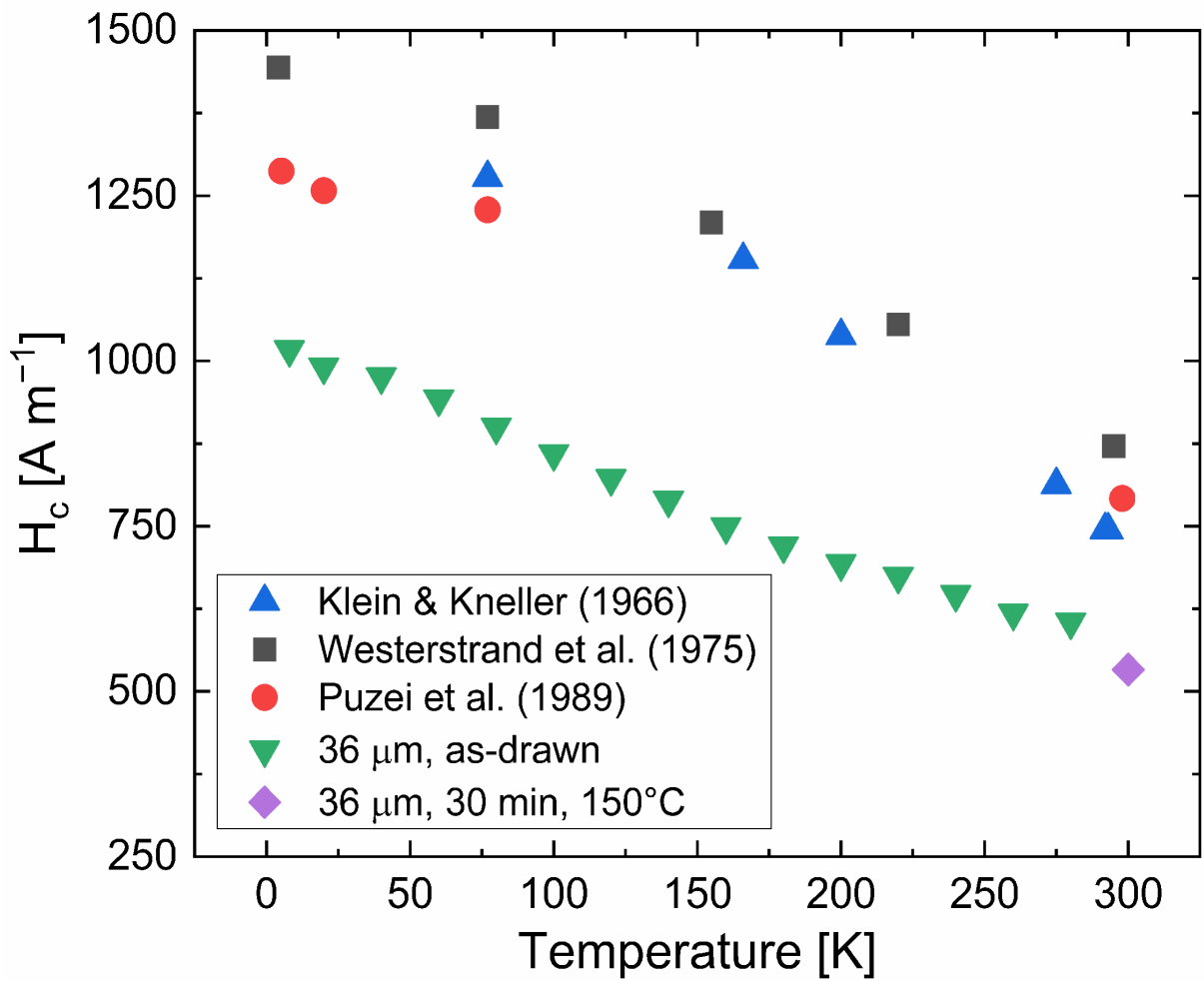

**Figure 4.** Measured and corrected temperature-dependent coercivities of a 36-μm wire as a function of temperature. Results are compared with coercivities calculated according to literature values of temperature dependent magneto-crystalline anisotropy $K_1$ [23–25]. The slightly decreased coercivity of the stress-relieved 36-μm wire is also included.

The slightly smaller value of coercivity after the annealing treatment of 150 °C for 30 min is also shown in Figure 4. The decrease in coercivity can be explained by decreasing the residual stresses imparted by the drawing process. Residual stresses couple with the magnetostriction of the material and lead to magneto-elastic anisotropy, effectively increasing the magnetic anisotropy. While the pure bcc Fe phase was of advantage when considering the achievable values of saturation magnetization, it is of disadvantage from a magnetostrictive point of view. While e.g., Si can be added (FeCuNbSiB [9]) for reaching zero magnetostriction, pure Fe is linked to a non-zero magnetostrictive value.

Another aspect of the discussion of the coercivity of severely deformed wires is the already found trend of decreasing coercivity with increasing applied strain (up to ~3.5) [18]. Therein, a pearlitic steel containing 0.7% C was subjected to wire drawing and its room temperature coercivity was determined. Figure 5 shows a comparison of the results from [18] together with a linear fit and the measurements described in this work. The trend in decreasing coercivity with increasing strain is clear and it is explained in [18] by volume reduction of the cementite, fragmentation and rotation of the remaining cementite lamellae in combination with a low-coercivity state of cementite after severe deformation [27]. The reasons for the small deviation of the linear extrapolation to the measured coercivity for 24- and 36-µm wires are manifold. The applicability of the linear extrapolation up to these large degrees of deformations is hypothetical, the materials slightly differ in C-content and experienced different drawing steps towards the final accumulated strain. Furthermore, there is a difference in the distance of the cementite lamellae in the non-deformed state. While this distance is 67 nm for the patented wire with 540-µm diameter [7], it is slightly larger (70–90 nm) for the initial state described in [18]. Already starting with a slightly smaller distance of lamellae could lead to a diminution of coercivity at higher strains.

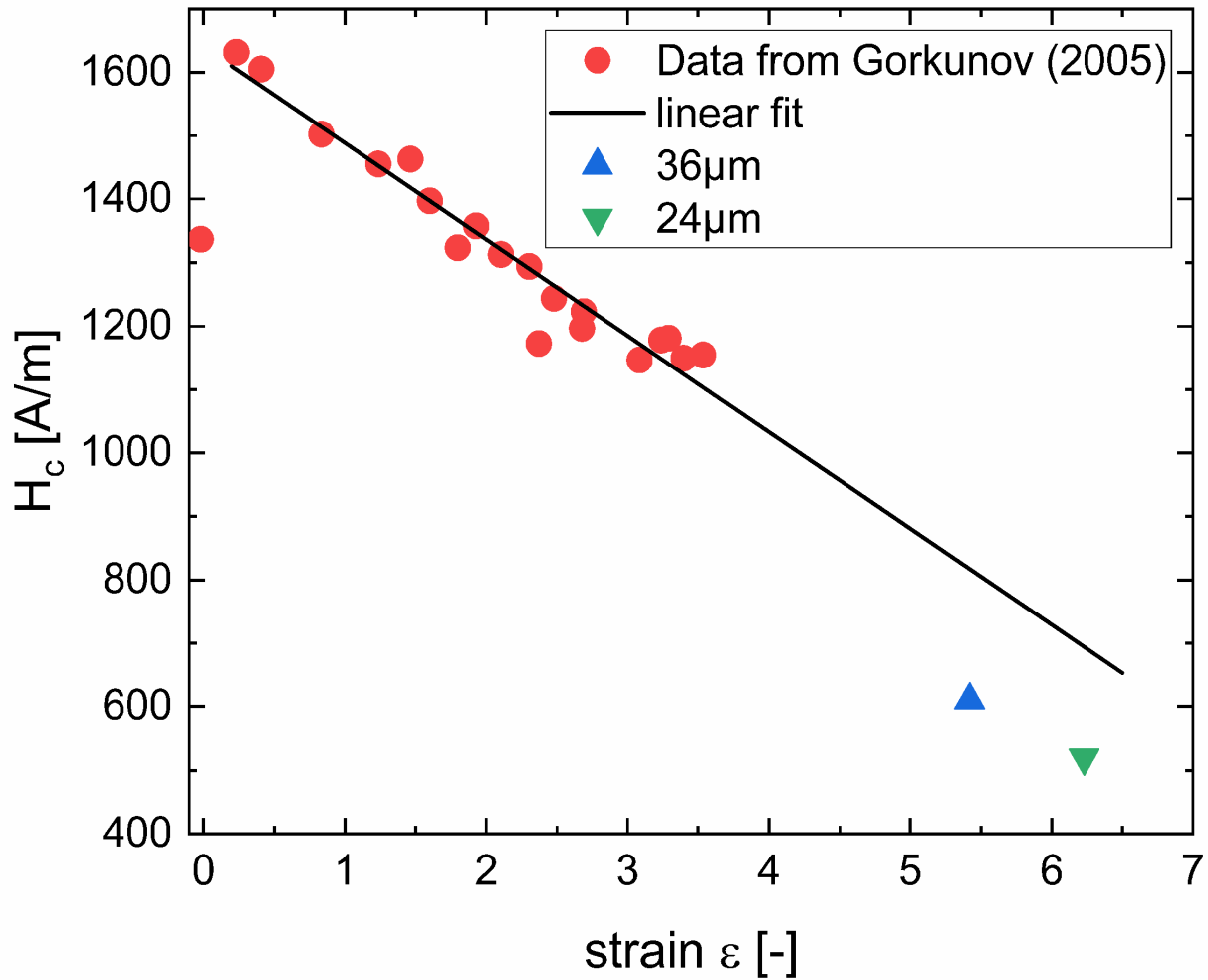

**Figure 5.** Data taken from [18], showing the coercivity determined on a steel containing slightly less C. This data is compared with results for even thinner wires, presented herein. The linear fit neglects the point at zero strain.

When comparing the two different approaches for describing the decreasing coercivity with increasing strain, it has to be mentioned that wire drawn materials feature a very pronounced texture. In [39], the pronounced fiber texture of bcc-W was presented. The texture prevails after annealing treatments, even in the recrystallized state. The development of the same texture was found for pearlitic steels experiencing strains smaller than 1.65 [40]. In conjunction with a pronounced texture, the crystallographic orientation of a grain correlates with adjacent grains or subgrains—counteracting the random anisotropy model, which relies per definition on a random orientation of crystallites. Summarizing, there are two ideas, which describe the found trend in coercivity. The first one is based on the random anisotropy model, while the second is based on microstructural changes, especially taking place for the cementite. Due to the presented arguments the one relying on microstructural changes for the cementite phase is more likely to be applicable.

## 4. Conclusions

Pearlitic wires - ferritic wires in different wire-drawn states - were subjected to magnetic measurements, focusing on the volume saturation magnetization and coercivity.

According to the indirect phase analysis based on the evaluation of saturation magnetization data, the drawn wires with diameters of 24 and 36 µm behave - from a magnetic point of view - like pure Fe with cementite being dissolved. This finding is supported by an independent measurement of the temperature dependent coercivity, also fitting to the behavior of pure Fe.

The measured values of volume saturation magnetization of the thin wires, who experienced a drawing strain larger than 5.4, are slightly higher as it can even be expected for pure Fe. This can be attributed to Fe being supersaturated by C, as it enlarges the magnetic moment of Fe-atoms.

It was demonstrated that an annealing treatment has a positive effect to magnetically soften the wire. Annealing at 150 °C for 30 min slightly reduces the coercivity, which is most likely due to a reduction of residual stresses being present from the severe deformation process.

**Author Contributions:** Conceptualization, R.P. and A.B.; methodology, S.W., M.S. and H.K.; formal analysis, S.W.; investigation, S.W., M.S. and L.W.; resources, H.K. and A.B.; data curation, S.W.; writing—original draft preparation, S.W.; writing—review and editing, S.W., M.S., L.W., H.K., A.H., R.P. and A.B.; visualization, S.W.; supervision, A.B.; project administration, A.B.; funding acquisition, A.B. All authors have read and agreed to the published version of the manuscript.

**Funding:** This project received funding from the European Research Council (ERC) under the European Union's Horizon 2020 research and innovation programme (Grant No. 757333).

**Acknowledgments:** The authors acknowledge the help of Shoji Goto to this publication in providing the materials for these investigations and Patrice Kreiml for his help in LEXT measurements.

**References**

1. Raabe, D.; Choi, P.-P.; Li, Y.; Kostka, A.; Sauvage, X.; Lecouturier, F.; Hono, K.; Kirchheim, R.; Pippan, R.; Embury, D. Metallic Composites Processed via Extreme Deformation: Toward the Limits of Strength in Bulk Materials. MRS Bull. 2010, 35, 982–991.

2. Hono, K.; Ohnuma, M.; Murayama, M.; Nishida, S.; Yoshie, A.; Takahashi, T. Cementite Decomposition in Heavily Drawn Pearlite Steel Wire. Scr. Mater. 2001, 44, 977–983.
3. Li, Y.J.; Choi, P.; Goto, S.; Borchers, C.; Raabe, D.; Kirchheim, R. Evolution of Strength and Microstructure during Annealing of Heavily Cold-Drawn 6.3 GPa Hypereutectoid Pearlitic Steel Wire. Acta Mater. 2012, 60, 4005–4016.
4. Herbig, M.; Raabe, D.; Li, Y.J.; Choi, P.; Zaefferer, S.; Goto, S. Atomic-Scale Quantification of Grain Boundary Segregation in Nanocrystalline Material. Phys. Rev. Lett. 2014, 112, 126103.
5. Li, Y.J.; Choi, P.; Borchers, C.; Westerkamp, S.; Goto, S.; Raabe, D.; Kirchheim, R. Atomic-Scale Mechanisms of Deformation-Induced Cementite Decomposition in Pearlite. Acta Mater. 2011, 59, 3965–3977.
6. Li, Y.J.; Choi, P.; Borchers, C.; Chen, Y.Z.; Goto, S.; Raabe, D.; Kirchheim, R. Atom Probe Tomography Characterization of Heavily Cold Drawn Pearlitic Steel Wire. Ultramicroscopy 2011, 111, 628–632.
7. Li, Y.; Raabe, D.; Herbig, M.; Choi, P.-P.; Goto, S.; Kostka, A.; Yarita, H.; Borchers, C.; Kirchheim, R. Segregation Stabilizes Nanocrystalline Bulk Steel with Near Theoretical Strength. Phys. Rev. Lett. 2014, 113, 106104.
8. Herzer, G. Grain Size Dependence of Coercivity and Permeability in Nanocrystalline Ferromagnets. IEEE Trans. Magn. 1990, 26, 1397–1402.
9. Herzer, G. Soft Magnetic Nanocrystalline Materials. Scr. Metall. Mater. 1995, 33, 1741–1756.
10. Kayser, F.X.; Litwinchuk, A.; Stowe, G.L. The Densities of High-Purity Iron-Carbon Alloys in the Spheroidized Condition. MTA 1975, 6, 55–58.
11. Topolovec, S.; Krenn, H.; Würschum, R. Electrochemical Cell for In Situ Electrodeposition of Magnetic Thin Films in a Superconducting Quantum Interference Device Magnetometer. Rev. Sci. Instrum. 2015, 86, 063903.
12. Osborn, J.A. Demagnetizing Factors of the General Ellpsoid. Phys. Rev. 1945, 67, 351–357.
13. Hohenwarter, A.; Völker, B.; Kapp, M.W.; Li, Y.; Goto, S.; Raabe, D.; Pippan, R. Ultra-Strong and Damage Tolerant Metallic Bulk Materials: A Lesson from Nanostructured Pearlitic Steel Wires. Sci. Rep. 2016, 6, 33228.
14. Neccas, D.; Klapetek, P. Gwyddion: An Open-Source Software for SPM Data Analysis. Cent. Eur. J. Phys. 2012, 10, 181–188.
15. Coey, J.M.D. Magnetism and Magnetic Materials; Cambridge University Press: Cambridge, UK, 2010; ISBN 978-0-511-67743-4.
16. Duman, E.; Acet, M.; Hülser, T.; Wassermann, E.F.; Rellinghaus, B.; Itié, J.P.; Munsch, P. Large Spontaneous Magnetostrictive Softening below the Curie Temperature of $Fe_3C$ Invar Particles. J. Appl. Phys. 2004, 96, 5668–5672.
17. Dick, A.; Körmann, F.; Hickel, T.; Neugebauer, J. Ab Initio Based Determination of Thermodynamic Properties of Cementite Including Vibronic, Magnetic, and Electronic Excitations. Phys. Rev. B 2011, 84, 125101.
18. Gorkunov, E.S.; Grachev, S.V.; Smirnov, S.V.; Somova, V.M.; Zadvorkin, S.M.; Kar'kina, L.E. Relation of Physical-Mechanical Properties to the Structural Condition of Severely Deformed Patented Carbon Steels at Drawing. Russ. J. Nondestruct. Test. 2005, 41, 65–79.

19. Medvedeva, N.I.; Kar'kina, L.E.; Ivanovskii, A.L. Electronic Structure and Magnetic Properties of the alpha- and gamma-Phases of Iron, Their Solutions with Carbon, and Cementite. Phys. Met. Metallogr. 2006, 101, 440.
20. Cadeville, M.C.; Lerner, C.; Friedt, J.M. Electronic Structure of Interstitial Carbon in Ferromagnetic Transition Metals Prepared by Splat-Quenching. Physica B+C 1977, 86–88, 432–434.
21. Suzuki, K.; Parsons, R.; Zang, B.; Onodera, K.; Kishimoto, H.; Shoji, T.; Kato, A. Nanocrystalline Soft Magnetic Materials from Binary Alloy Precursors with High Saturation Magnetization. AIP Adv. 2019, 9, 035311.
22. Hono, K.; Sepehri-Amin, H. Strategy for High-Coercivity Nd–Fe–B Magnets. Scr. Mater. 2012, 67, 530–535.
23. Klein, H.-P.; Kneller, E. Variation of Magnetocrystalline Anisotropy of Iron with Field and Temperature. Phys. Rev. 1966, 144, 372–374.
24. Westerstrand, B.; Nordblad, P.; Nordborg, L. The Magnetocrystalline Anisotropy Constants of Iron and Iron-Silicon Alloys. Phys. Scr. 1975, 11, 383–386.
25. Puzei, I.M.; Sadchikov, V.V. Dependence of the Magnetic-Anisotropy Energy in Iron on the Magnetic Field. Sov. Phys. JETP 1990, 70, 137–139.
26. Herzer, G. The Random Anisotropy Model. In Properties and Applications of Nanocrystalline Alloys from Amorphous Precursors; Idzikowski, B., Švec, P., Miglierini, M., Eds.; Springer: Berlin/Heidelberg, Germany, 2005; Volume 184, pp. 15–34. ISBN 978-1-4020-2963-9.
27. Ul'yanov, A.I.; Elsukov, E.P.; Chulkina, A.A.; Zagainov, A.V.; Arsent'eva, N.B.; Konygin, G.N.; Novikov, V.F.; Isakov, V.V. The Role of Cementite in the Formation of Magnetic Hysteresis Properties of Plastically Deformed High-Carbon Steels: I. Magnetic Properties and Structural State of Cementite. Russ. J. Nondestruct. Test. 2006, 42, 452–459.
28. Chulkina, A.A.; Ul'yanov, A.I.; Arsent'eva, N.B.; Zagainov, A.V.; Gorkunov, E.S.; Zadvorkin, S.M.; Somova, V.M. The Role of Cementite in the Formation of Magnetic Hysteresis Properties of Plastically Deformed High-Carbon Steels: II. Magnetic Properties of Patented Wire Made of Steel 70. Russ. J. Nondestruct. Test. 2006, 42, 460–467.
29. Chulkina, A.A.; Ul'yanov, A.I.; Gorkunov, E.S. The Role of Cementite in the Formation of Magnetic Hysteresis Properties of Plastically Deformed High-Carbon Steels: III. Magnetic Properties of PatentedWire Made of Steel 25. Russ. J. Nondestruct. Test. 2008, 44, 309–317.
30. English, A.T. Influence of Temperature and Microstructure on Coercive Force of 0.8% Steel. Acta Met. 1967, 15, 1573–1580.
31. Stückler, M.; Weissitsch, L.; Wurster, S.; Felfer, P.; Krenn, H.; Pippan, R.; Bachmaier, A. Magnetic Dilution by Severe Plastic Deformation. AIP Adv. 2020, 10, 015210.
32. Li, Y.J.; Kostka, A.; Choi, P.; Goto, S.; Ponge, D.; Kirchheim, R.; Raabe, D. Mechanisms of Subgrain Coarsening and Its Effect on the Mechanical Properties of Carbon-Supersaturated Nanocrystalline Hypereutectoid Steel. Acta Mater. 2015, 84, 110–123.
33. Zhang, X.; Hansen, N.; Godfrey, A.; Huang, X. Dislocation-Based Plasticity and Strengthening Mechanisms in Sub-20 Nm Lamellar Structures in Pearlitic Steel Wire. Acta Mater. 2016, 114, 176–183.

34. Djaziri, S.; Li, Y.; Nematollahi, G.A.; Grabowski, B.; Goto, S.; Kirchlechner, C.; Kostka, A.; Doyle, S.; Neugebauer, J.; Raabe, D.; et al. Deformation-Induced Martensite: A New Paradigm for Exceptional Steels. Adv. Mater. 2016, 28, 7753–7757.
35. Taniyama, A.; Takayama, T.; Arai, M.; Hamada, T. Structure Analysis of Ferrite in Deformed Pearlitic Steel by Means of X-ray Diffraction Method with Synchrotron Radiation. Scr. Mater. 2004, 51, 53–58.
36. Liu, J.; Schliep, K.; He, S.-H.; Ma, B.; Jing, Y.; Flannigan, D.J.;Wang, J.-P. Iron Nanoparticles with Tunable Tetragonal Structure and Magnetic Properties. Phys. Rev. Mater. 2018, 2, 054415.
37. Using SQUID VSM Superconducting Magnets at Low Fields, Application Note 1500-011; Quantum Design: Tokyo, Japan, 2010.
38. Yamamoto, S.; Terai, T.; Fukuda, T.; Sato, K.; Kakeshita, T.; Horii, S.; Ito, M.; Yonemura, M. Magnetocrystalline Anisotropy of Cementite Pseudo Single Crystal Fabricated under a Rotating Magnetic Field. J. Magn. Magn. Mater. 2018, 451, 1–4.
39. Nikolic, V.; Riesch, J.; Pippan, R. The Effect of Heat Treatments on Pure and Potassium Doped Drawn Tungsten Wires: Part I—Microstructural Characterization. Mater. Sci. Eng. A 2018, 737, 422–433.
40. Guo, N.; Luan, B.; Wang, B.; Liu, Q. Microstructure and Texture Evolution in Fully Pearlitic Steel during Wire Drawing. Sci. China Technol. Sci. 2013, 56, 1139–1146.